\documentclass[aps,showkeys,superscriptaddress,twocolumn,prb,longbibliography]{revtex4-1}
\usepackage{graphicx}% Include figure files
\usepackage{dcolumn}% Align table columns on decimal point
\usepackage{multirow}
\usepackage{booktabs}
\usepackage{bm,color}
\usepackage{braket}
\usepackage{amsmath,amssymb}
\usepackage[colorlinks,linkcolor=blue,citecolor=blue,urlcolor=blue,hyperindex,CJKbookmarks]{hyperref}
\usepackage{epstopdf}

\begin {document}

\title{\bf  First principles calculation of shift current in chalcopyrite semiconductor ZnSnP$_2$}
\author{Banasree Sadhukhan}
 \email{b.sadhukhan@ifw-dresden.de}
 \affiliation{Leibniz Institute for Solid State and Materials Research IFW Dresden, Helmholtzstr. 20, 01069 Dresden, Germany}
\author{Yang Zhang}
\affiliation{Leibniz Institute for Solid State and Materials Research IFW Dresden, Helmholtzstr. 20, 01069 Dresden, Germany}
\affiliation{Department  of  Physics,  Massachusetts  Institute  of  Technology,  Cambridge,  Massachusetts  02139,  USA}
\affiliation{Max Planck Institute for Chemical Physics of Solids, 01187 Dresden, Germany}
\author{Rajyavardhan Ray}
 \email{r.ray@ifw-dresden.de}
 \affiliation{Leibniz Institute for Solid State and Materials Research IFW Dresden, Helmholtzstr. 20, 01069 Dresden, Germany}
\affiliation{Dresden Center for Computational Materials Science, TU Dresden, D-01062 Dresden, Germany}
\author{Jeroen van den Brink}
 \affiliation{Leibniz Institute for Solid State and Materials Research IFW Dresden, Helmholtzstr. 20, 01069 Dresden, Germany}
 \affiliation{Dresden Center for Computational Materials Science, TU Dresden, D-01062 Dresden, Germany}

\date{\today}

\begin{abstract}
The bulk photovoltaic effect generates intrinsic photocurrents in
materials without inversion symmetry. Shift current is one
of the bulk photovoltaic phenomena related to the Berry phase of the
constituting electronic bands: photo-excited carriers coherently
shift in real space due to the difference in the Berry connection
between the valence and conduction bands. Ferroelectric
semiconductors and Weyl semimetals are known to exhibit such
nonlinear optical phenomena. Here we consider chalcopyrite
semiconductor ZnSnP$_2$ which lacks inversion symmetry and calculate
the shift current conductivity. We find that the magnitude of the
shift current is comparable to the recently measured values on other
ferroelectric semiconductors and an order of magnitude larger than
bismuth ferrite. The peak response for both optical and shift
current conductivity, which mainly comes from P-3$p$ and Sn-5$p$
orbitals, is several eV above the bandgap. 
\end{abstract}

\maketitle

%%%%%%%%%%%%%%%%%%%%%%%%%%%%%%%%%%%%%%%%%%%%%%%%%%%%%%%%%%%%%%%%%%%%%%%%%%%%%%%%%%%%%%%%%%%%%%%%%%%%%%%%%%%

\section {Introduction}
Ternary compounds A$^{II}$B$^{IV}${C}$^V_2$ and
A$^I$B$^{III}${C}$^{VI}_2$ (where, A, B = metals, C = sulfur/nitrogen family) 
having chalcopyrite structure are of
considerable interest because of their structural, mechanical,
thermoelectric and nonlinear optical properties \cite{pandey1998}. They are also promising
materials for spintronics application because of the ability to host
ferromagnetism at room temperature \cite{medvedkin2000,cho2002}. The
chalcopyrite structures are derived from the binary analogs
M$^{III}$C$^V$ and M$^{II}$C$^{VI}$ in cubic zinc-blende structures by
doubling the unit cell along $c$, leading to a body-centered tetragonal unit
cell.
Each cation (anion) is surrounded by four nearest-neighbor anions
(cations) as in the zinc-blende structure. 
The A and B cations alternatively occupy the Zn-positions and form a 
tetrahedral bonding of the two cation sublattice. 
The reduced symmetry lowers the band
gap significantly in the ternary compounds compared to their binary analogs
\cite{yeh1994}.
This spatial symmetry reduction also plays an important role to realize the
topological insulating and Weyl semimetallic phases in some ternary
chalcopyrites \cite{feng2011,ruan2016,juneja2018,lau2019,
fu2007,bernevig2006,xiao2010,chadov2010,lin2010}.

Of particular interest is the ternary 
compound ZnSnP$_2$ (ZSP), type A$^{II}$B$^{IV}${C}$^V_2$, which 
is now recognized as an alternative photoabsorber
material in solar cell applications \cite{scanlon2012,kumagai2014}. It undergoes a
structural transition from the ordered chalcopyrite ZnSnP$_2$ (CH-ZSP)
structure to a disordered sphalerite structure (SP-ZSP) at 990 K
\citep{shay2017}. In
SP-ZSP, the Sn and Zn atoms are randomly distributed over the cation
sub-lattice. In comparison, in the ordered CH-ZSP, the P$^{3-}$ anions
are surrounded by two Zn$^{2+}$ (A-type) and two Sn$^{4+}$ (B-type) cations while each cation is
surrounded by four anions. Due to the possibility of bandgap
engineering and tunability of the electronic and optical properties, 
the electronic structure and properties of ZSP  
are being investigated both theoretically and experimentally
\cite{nakatani2008, st-jean2010,
scanlon2012, mishra2013, sahin2012,xu2015,mukherjee2018}.
In comparison to other well known chalcopyrite ternary compounds, an
important feature of CH-ZSP is that the ground state is a trivial insulator,
but lacks inversion symmetry due to the displacement of the anion
positions (anion-shift) towards one of the cation as compared to the
ideal case obtained by doubling the cubic zinc-blende unit cell (see Sec.
\ref{sec:results} for details). It is also
anisotropic due to the presence of two types of cationic bonding which
also gives rise to high birefringence.

An interesting and potentially useful property of  non-centrosymmetric
crystals is that in such materials symmetry allows incident photons to
induce a photocurrent. This is a bulk photovoltaic phenomenon and the
induced current is referred to as the shift current \cite{sipe2000,
morimoto2016_prb, nastos2006,nastos2010,zhang2018_rc,morimoto2016_scadv}. 
In contrast to a conventional drift photocurrent under an electric field,
the shift current originates from the charge center shifts in real
space due to difference in the Berry connection between the
valence and conduction bands involved in the optical excitation process
\cite{morimoto2016_prb, morimoto2016_scadv}. Recently, Weyl semimetals
(WSMs) have been theoretically investigated for such nonlinear optical
phenomena \cite{zhang2018_rc,goswami2015,ishizuka2016,sodemann2015,
morimoto2016_prb2, konig2017,zhang2018, yang2017}.

The shift current with a less
dissipative character has remarkable advantages over the conventional
drift photocurrent driven by a built-in potential or external electric
field \citep{choi2009,yang2010,grinberg2013}. For example, it depends on
the polarization direction of the incident photon field, is insensitive to the sample
resistivity or barrier formation near the electrodes \cite{ogawa2017}, and is
also independent of the external bias voltage \cite{tan2016, young2012}.

Moreover, photocurrents induced by optical transitions obeying dipole and
polarization selection rules naturally permit ultrafast manipulation. In
particular, shift currents induced by a properly tuned external pulsed
photon sources can create coherent electromagnetic wave emission in the
terahertz frequency regime, where control of the ellipticity and
chirality over a broad spectral range is notoriously
difficult~\cite{amer2005,gao2020}.

Here we show, using a recently developed multi-band approach,
\cite{zhang2018_rc} 
that the shift current conductivity \cite{zhang2018_rc, baltz1981, kraut1979} in CH-ZSP is comparable to
that in SbSI \cite{sotome2018}, and an order of magnitude larger than the famous multiferroic
BiFeO$_3$ \cite{young2012_bfo}. 
The multi-band approach involves the full set
of Bloch states and a sum over all intermediate states participating in
three-band transitions is considered.
The three-band virtual transitions make the dominant contributions and are
distributed uniformly in the momentum space. Naturally, in comparison to
the widely used two-band effective models, estimates based on the
multi-band approach are accurate and highly desirable for materials application. 

A key challenge in an accurate density functional theory (DFT) based
description of insulating materials is the well known problem of bandgap
underestimation by the local and semi-local functionals. 
These problems can be cured by employing schemes which take into account
the self-energy of a many-body electronic system, such as the GW
approximation \cite{aryasetiawan1998}, and the hybrid exchange correlation
(HSE) functional \cite{heyd2003}. However, they are computationally very
expensive. 
At the same time, while the HSE functionals improve the bandgap close
to the experimental value, it may overestimate the lattice constants and the
atomic displacement associated with structural distortion which may
eventually lead to inaccurate estimate of material properties. For example,
in the non-centrosymmetric compound BaTiO$_3$, such overestimation affects
the ferroelectricity and gives inaccurate optical response 
 \cite{sanna2011, gupta2004, bagayoko1998}. 

Traditionally, the deficiency associated with the bandgap is addressed by using
a simple ``scissors operation" \cite{godby1988, allan1989} on the
standard DFT [using generalized gradient approximation/local density approximation (GGA/LDA)] bands whereby the conduction bands
are rigidly shifted such that the resulting electronic bandgap matches
with the experimental value. Within this procedure, the optical response
obtained within LDA/GGA is shifted by the same amount (referred to as scissor-shift
in the following) and retains the features obtained from standard
DFT \cite{nastos2005}.  As an
alternative, a semi-empirical DFT+$U$ approximation might be used to
improve the bandgap values. Very recently, an empirical Tran-Blaha
modified Becke-Johnson (TB-mBJ) potential \cite{mbj2009} was shown to lead to an accuracy
comparable to the much expensive hybrid functional and GW approximation
at a computational cost comparable to standard DFT calculations.
Here, we thus, consider the latter three approaches, {\it viz.}, the
scissors operation (GGA$+\Delta$), DFT+$U$ and TB-mBJ methods and discuss their
implications for the electronic and optical properties of CH-ZSP.

%%%%%%%%%%%%%%%%%%%%%%%%%%%%%%%%%%%%%%%%%%%%%%%%%%%%%%%%%%%%%%%%%%%%%%%%%%%%%%%%%%%%%%%%%%%%%%%%%%%%%%%%%%% 
 
\section{Computational details}

\par We performed density functional theory (DFT) calculations within
the Perdew-Burke-Ernzerhof (PBE) implementation \cite{pbe1996} of 
the GGA  functional using the full-potential
local-orbital (FPLO) code \cite{klaus1999,fplo_web}. Self-consistent
calculations employing the default scalar relativistic approximation were performed on
a $k$-mesh with 18 $\times$ 18 $\times$ 18 subdivisions. 
Starting from the experimental structure \cite{vaipolin1968}, several crystal
structures with different unit cell volumes $V$ were considered: $ 0.90
V_{\rm exp} \leq V \leq 1.10 V_{\rm exp}$, where $V_{\rm exp}$ is the
unit cell volume of the experimental crystal structure. For each case, the internal
parameters (atomic positions) were optimized such that net force on each
atom was less than 1 meV/{\AA} and the ground state
energy was evaluated. The optimized structure was considered for
further detailed study of electronic and optical properties. 
The spin-orbit effects are expected to be small and were, therefore,
not considered  (see Appendix
\ref{sec:app_a}).

To overcome the issue of bandgap underestimation, both DFT+$U$ 
and TB-mBJ calculations were carried out. The on-site orbital
dependent electron electron correlations ($U$) were applied to Zn-3$d$
as well as P-3$p$ states and the evolution of the bandgap was studied. 

The TB-mBJ calculations \cite{mbj2009} were carried out using the full-potential 
Augmented Plane Waves + local orbital (APW+lo) method as implemented in the WIEN2k code
\cite{wien2k}. A good quantitative and qualitative agreement between
the two codes were obtained within the scalar relativistic GGA
calculations. For the TB-mBJ potential, the self-consistent $c$ parameter
was used \cite{mbj2009}. The energy convergence of the obtained solutions is
better than $10^{-5}$ Ryd per unit cell and the charge convergence is
better than $10^{-4}$ e/a.u.$^3$.

The optical properties within the linear response theory were obtained
using the well-known relations: 
the imaginary part of the dielectric function is given by 
\begin{align}
    \epsilon_2^{\alpha \beta}(\omega) = {\rm Im} [\epsilon_{\alpha \beta}(\omega)] 
        &=-\frac{4\pi^2 e^2} {m_0^2 \omega^2} \int d{\bf k} \sum_{n,l} \left(f_n-f_l \right)  \nonumber \\
        & \times
    \frac{   \langle \mathbf{k}n | \hat{v}_{\alpha} | \mathbf{k}l
    \rangle  \langle \mathbf{k}l | \hat{v}_{\beta} | \mathbf{k}n \rangle }
                {  (E_{\mathbf{k}n}-E_{\mathbf{k}l}-\hbar \omega -
                i\delta) }\,,
\end{align}
where, $\alpha,\beta = (x,y,z)$ are the Cartesian coordinates,
$\hat{v}_{\alpha}=\hat{p}_{\alpha}/m_0$ is the velocity operator along
$\alpha$, $m_0$ is the free electron mass, $|{\bf k}n
\rangle$ are the wavefunction corresponding to the band with energy $E_{{\bf k}n}$ at
momentum ${\bf k}$ and index $n$, $f_n \equiv f(E_{\mathbf{k}n})$ is the Fermi function for the
state with energy $E_{\mathbf{k}n}$, and $\hbar \omega$ is the
incident photon energy. $\delta=\hbar/\tau_s$ is
the broadening parameter which depends inversely on the single particle
relaxation time associated with the quantum mechanical broadening
$\tau_s$.
The real part can be obtained via
the Kramer-Kronig relation:
\begin{equation}
    \epsilon_1^{\alpha \beta}(\omega) = {\rm Re} {[\epsilon_{\alpha \beta}(\omega)]} = \delta_{\alpha \beta}
    + \frac{1}{\pi} \mathcal{P}\int_{-\infty}^\infty \ d\omega'\
    \frac{{\rm Im} [\epsilon_{\alpha \beta}(\omega')]}{\omega -\omega'} \,.
\end{equation}
All optical response functions can now be derived from these. In particular, the optical conductivity is 
\begin{equation}
    \sigma_{\alpha \beta}(\omega) = \frac{\omega\epsilon_2^{\alpha \beta}(\omega)}{4\pi} \,.
\end{equation}

To calculate the shift current response, we used the general relation
for the photoconductivity in quadratic response theory
\cite{zhang2018_rc,baltz1981, kraut1979}:
\begin{equation}
\begin{aligned}
    \sigma^{\gamma}_{\alpha \beta} &= \frac{|e|^3}{8\pi^3 \omega^2} {\rm Re} \bigg\{ \phi_{\alpha \beta}
                        \sum_{\Omega=\pm \omega} \sum_{l,m,n} \int_{BZ} d{\bf k} (f_l- f_n) \\
                        & \times \frac{ \langle {\bf k}n | \hat{v}_\alpha   | {\bf k}l \rangle 
                    \langle {\bf k}l |  \hat{v}_\beta   | {\bf k}m \rangle
                    \langle {\bf k}m |  \hat{v}_\gamma   | {\bf k}n \rangle}
            {(E_{\mathbf{k}n}-E_{\mathbf{k}m}-i\delta)(E_{\mathbf{k}n}-E_{\mathbf{k}l} + \hbar \Omega- i\delta)} \bigg\}.
\end{aligned}
\label{shift-eq}
\end{equation}
The conductivity $\sigma_{\alpha \beta}^{\gamma}$ ($\alpha,\beta,\gamma = x,y,z$) is a third rank
tensor representing the photocurrent $J_\gamma$ generated by an electrical
field via $J_\gamma=\sigma_{\alpha \beta}^\gamma {{\mathcal
E}^*_\alpha}{\mathcal E}_\beta$.  $\phi_{\alpha \beta}$ is the phase difference between the
driving field ${\mathcal E}_\alpha$ and ${\mathcal E}_\beta$. The real
part of the integral in
Eq. (\ref{shift-eq}) describes the shift current response under
a linearly polarized light. Note that while the above equation doesn't
    explicitly reflect the topological nature of the shift current, it
depends on the topological Berry connection and Berry
curvature \cite{young2012,morimoto2016_prb,zhang2018_rc}.

The starting point for shift current calculation is a bandstructure and
the corresponding eigenstates and energies in the Brillouin zone. To
this end, a tight binding model was obtained using maximally projected Wannier
functions (WFs) for the Zn-3$d$, Sn-4$d$, 5$s$, 5$p$ and P-3$p$ orbitals
in the energy range of -7.0 eV to 5.0 eV. The typical mismatch between
the tight-binding model derived from such Wannier functions and the
self-consistent DFT bandstructure was $\lesssim 1$ meV. For
the integral in Eq. (\ref{shift-eq}), the Brillouin zone (BZ) was sampled
by a $200 \times 200 \times 200$ $k$-mesh with satisfactory
convergence. The value of the conductivity changes by less than $3-4$\%
above that $k$-mesh. 
A typical value of the broadening parameter was used for both linear and
non-linear response: in ordinary metals and semiconductors,
the ratio of the transport relaxation time ($\tau_t$) and the single particle relaxation time
associated with the quantum mechanical broadening ($\tau_s$) is
$\tau_t/\tau_s = 1 $ \cite{xhong2009, hwang2008, bdas1993,sadhukhan2017}. In
semiconductors, the transport relaxation
time $\tau_t \approx$ femtoseconds (10$^{-15}$ sec.) at room temperature
\cite{fukumoto2014, inuzuka2004}, leading to $\delta \approx
\hbar/\tau_s=0.1$ at room temperature.

ZnSnP$_2$ belongs to the $\it{D_{2d}}$
(-4$\it{m}$2) point group in the ferroelectric phase. Therefore, it has the second-order
photoconductivity ($\sigma^\gamma_{\alpha \beta}$)  tensor of the form 
\[
    \sigma_{\alpha \beta}^{\gamma}=
  \left( {\begin{array}{ccccccc}
   0 & 0 & 0 & \sigma^x_{yz} & 0 & 0 \\
   0 & 0 & 0 & 0 & \sigma^y_{xz} & 0\\
   0 & 0 & 0 & 0 & 0 & \sigma^z_{xy}\\
  \end{array} } \right)
\]
The second-harmonic susceptibility $\chi_{\alpha \beta}^{\gamma}$
($\chi_{\alpha \beta}^{\gamma}=\sigma_{\alpha \beta}^{\gamma}/2i\omega\epsilon$) is governed by the same
symmetry and, therefore, has similar form. The crystal has a mirror
reflection $M_{xy}$ in the $x-y$
plane, which exchanges $x$ and $y$ indexes. In addition, the $4_2$ screw
rotation symmetry about the $z$ axis gives $\sigma^x_{yz}=\sigma^y_{xz}$,
leaving only two independent nonlinear optical photoconductivity tensor
elements $\sigma{^x_{yz}}$ and $\sigma{^z_{xy}}$.

%%%%%%%%%%%%%%%%%%%%%%%%%%%%%%%%%%%%%%%%%%%%%%%%%%%%%%%%%%%%%%%%%%%%%%%%%%%%%%%%%%%%%%%%%%%%%%%%%%%%%%%%%%%

\section{Results and Discussion}
\label{sec:results}

\begin{figure}[b!]
\hskip 0.9cm\includegraphics[width=0.49\textwidth,angle=0]{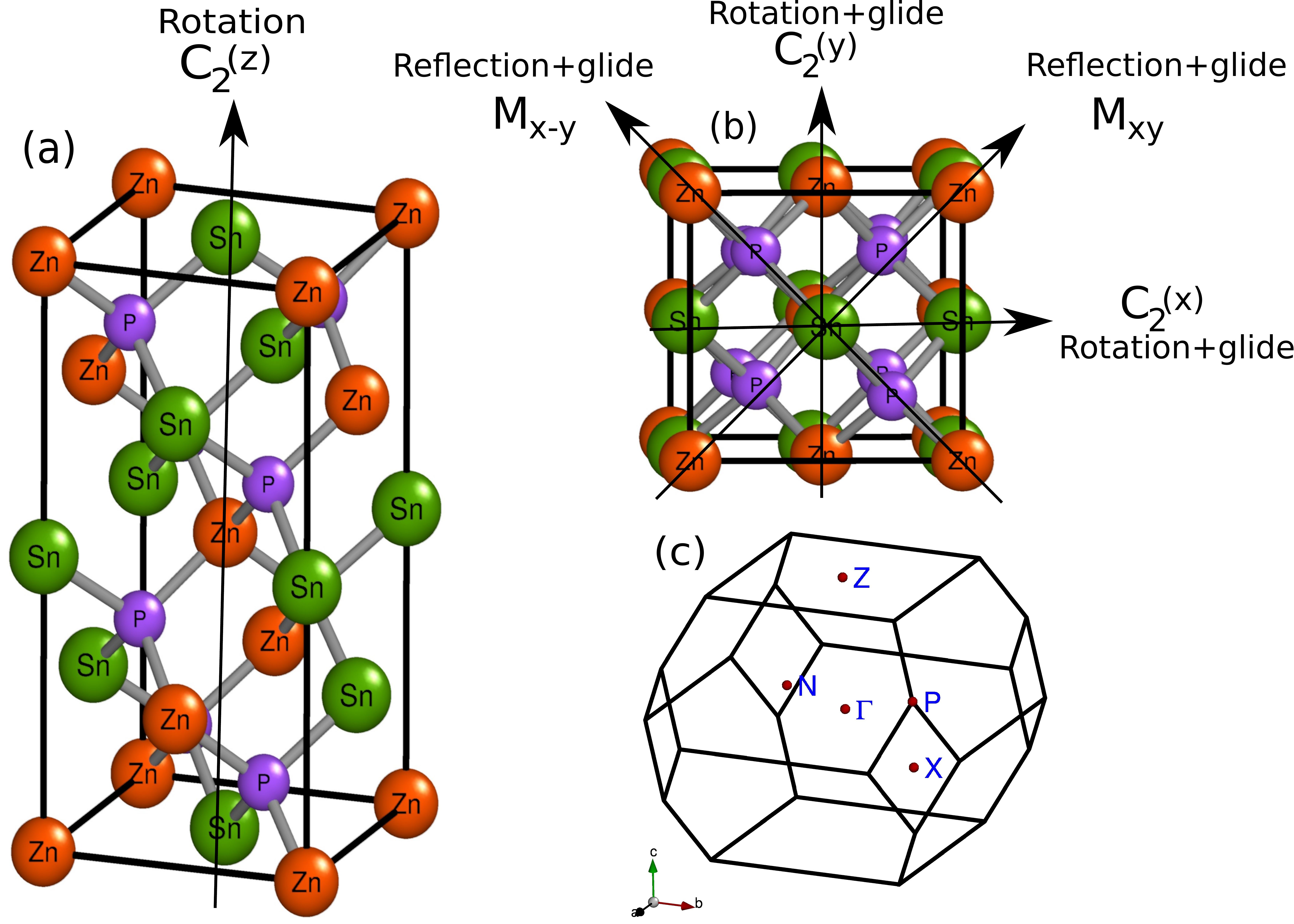}
\caption{Unit cell of the chalcopyrite ZnSnP$_2$ (CH-ZSP) lattice:
    (a) side view, (b) top view. It has two twofold glide rotational
and mirror symmetries C$_2$(x), C$_2$(y), M$_{xy}$ and M$_{x-y}$. (c)
Brillouin zone (BZ) along with the high symmetric points.
\label{unit_cell}}
\end{figure}

Ternary ZSP crystallizes in a body-centered
tetragonal structure which in the chalcopyrite phase (CH-ZSP) has the space group 
I$\bar{4}$2d (No. 122). It has eight atoms per primitive unit cell. Basically,
it is a superlattice of zinc-blende structure obtained by doubling the zinc-blende
unit cell along the $z$ direction. The unit cell of ZnSnP$_2$ is shown in
Fig. \ref{unit_cell}(a). 
In an
ideal zinc-blende structure of binary compound, each anion has four
similar cations as nearest neighbors. So, all the four bond lengths are
equal and the charge distribution is identical around each bond.
Consequently, in a binary compound with zinc-blende structure, $u$ is 0.25
and $\eta = c/a=1$. Therefore, the ideal case for the doubled unit
cell corresponds to $u=0.25$ and $\eta = c/a =
2$.
    
In CH-ZSP,
each anion has two Zn and two Sn cations as nearest
neighbor. Due to dissimilar atoms as neighbors the anion acquires an
equilibrium position closer to one pair of cation than to the other. The
displacement of the position of anion thus leads to bond alternation. In
the most general case, $u \neq$ 0.25 and $\eta$
$\neq$ 2. 
In contrast to other chalcopyrite
compounds, CH-ZSP lacks tetragonal distortion ($\eta =
2$) but
exhibits displacement of anions towards the smaller cation (anion
shift).
The positions of the different types of atoms in
the tetragonal unit cell are: Zn atom at (0, 0, 0); Sn atom at (0, 0,
0.5) and P atom at ($u$, 0.25, 0.125), where $u$ is
the anion displacement parameter. 
The equilibrium lattice parameters and the optimal internal parameters for
CH-ZSP, along with the corresponding experimental values, are presented in Table \ref{tab1}.
Compared to the binary compound, 
the cubic symmetry is broken and the
non-centrosymmetric CH-ZSP crystal has two twofold glide rotational
symmetries $C_2(x)$, $C_2(y)$, and two glide mirror symmetries $M_{xy}$
and $M_{x-y}$. It also has a twofold rotational symmetry along $z$ [see
Fig. \ref{unit_cell}]. 

\begin{table}[t]
    \small
    \caption{The experimental and equilibrium structural parameters for
        CH-ZSP. Please see text for details. The experimental parameters
    were taken from Ref. [\onlinecite{vaipolin1968}].}
    \begin{tabular*}{0.39\textwidth}{ p{2.7cm} p{1.3 cm} p {1.3 cm} p{1.3 cm}}
    \hline\hline
      ZnSnP$_2$  & $a$ (\AA)  & $c$ (\AA)  & $u$  \\
    \hline
	    Experimental  & 5.7382 & 11.4764 & 0.239  \\
	    Theoretical & 5.7382 & 11.4764 &  0.2272   \\
    \hline
    \end{tabular*}
    \label{tab1}
\end{table}

\begin{figure}[h!]
\includegraphics[width=0.41\textwidth,angle=0]{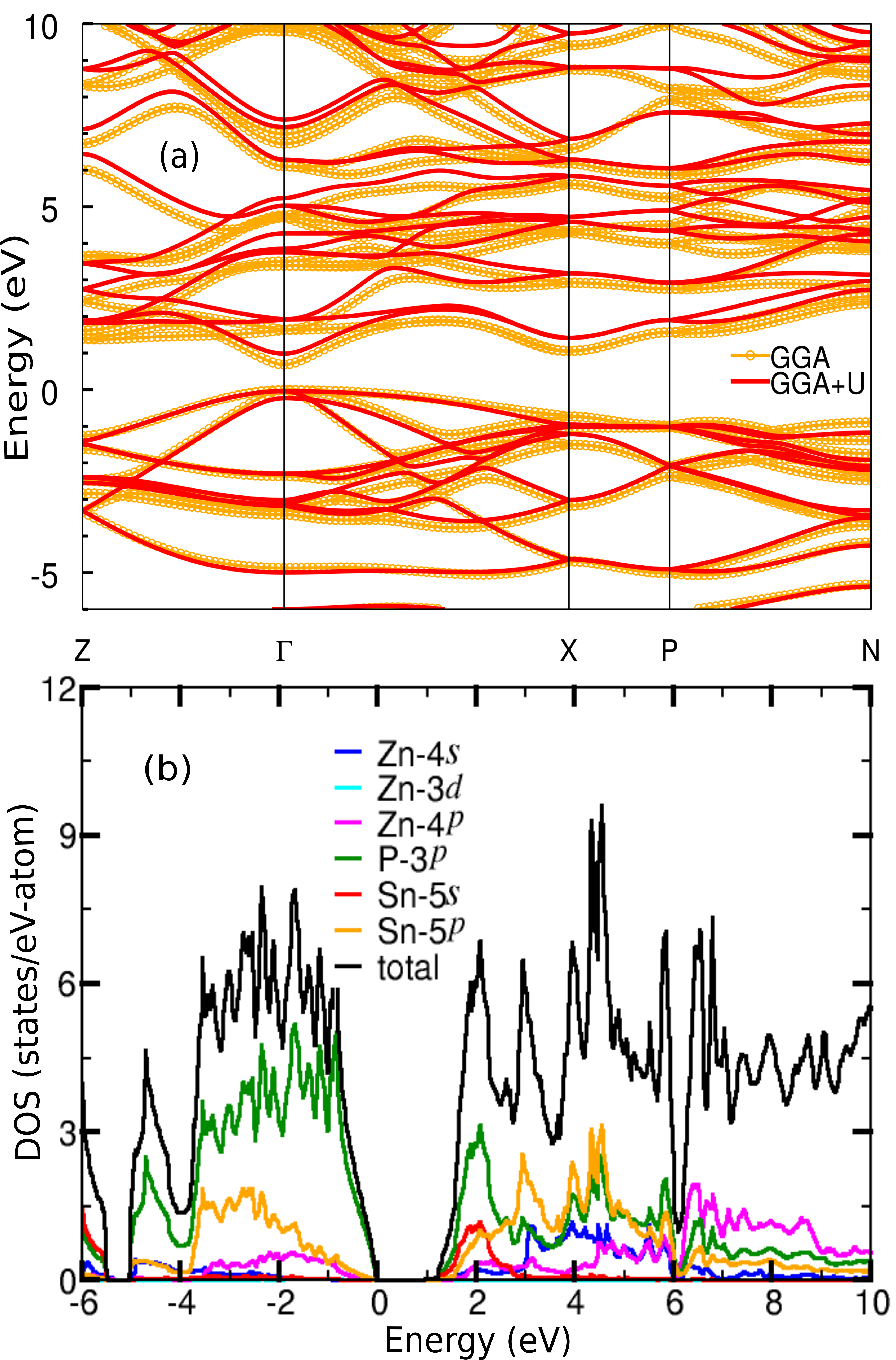}\\
\includegraphics[width=0.45\textwidth,angle=0]{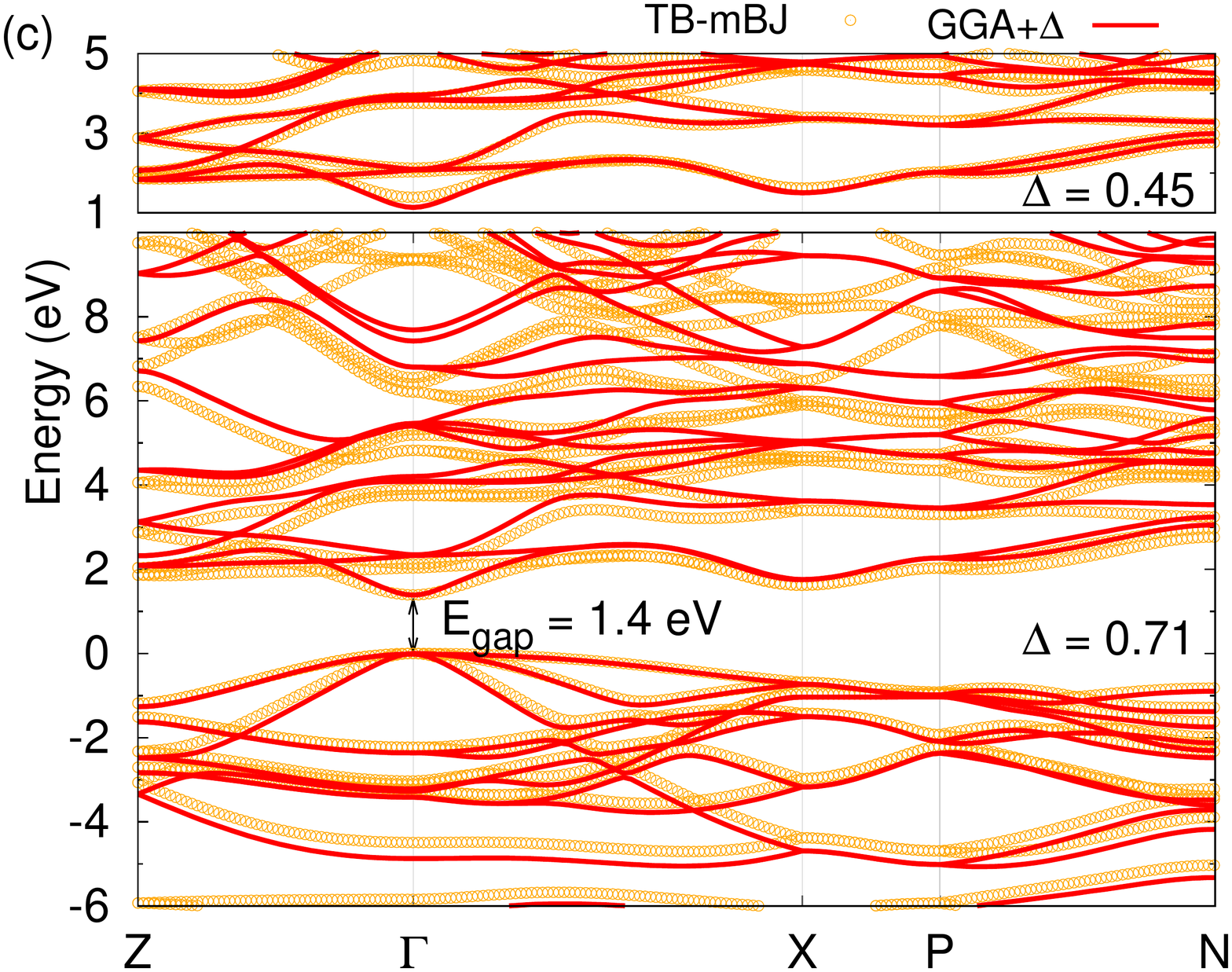}
\caption{(a) The bandstructure from GGA and GGA+$U$,
    showing a direct bandgap at $\Gamma$. The GGA band
    gap is $\sim 0.69$ eV whereas the GGA+$U$ bandgap is $\sim 1.05$
    eV, obtained with $U_d = 10$ eV and  $U_p = 2$ eV (see text for
    details). (b)
    The total and partial density of states within GGA+$U$.  (c) Comparison of
    bandstructures obtained within GGA$+\Delta$ and TB-mBJ calculations. For
    comparison, the GGA conduction bands have been scissor-shifted by $\Delta =
    0.71$ eV (bottom panel) to match the bandgap, and $\Delta=0.45$ eV
     in an energy range of 1 eV and 5 eV (top panel), showing the
     qualitative and quantitative
    agreement between the two methods in the description of the
    conduction band states.
}
\label{band_dos} 
\end{figure}

The bandstructure along the high symmetry lines in the Brillouin zone
and the density of states (DOS) is shown in Fig. \ref{band_dos}.
A direct bandgap is found at the $\Gamma$ point (see Fig. \ref{band_dos}(a)). 
Within GGA, the bandgap is $\sim 0.69$ eV
in good agreement with earlier calculations \cite{xu2015}. 
This is, however, merely $\sim 41\%$ of the experimental gap of 1.68
eV \cite{st-jean2010}, implying that a scissor-shift of $\Delta \sim
    1$ eV should be applied to obtain a quantitative agreement with the
experimental results.

Within the $+U$ scheme,  the bandgap can reach 1.05 eV upon adjusting the $U$ parameter. Since the dominant
contribution across the Fermi energy is due to P-3$p$ states [see Fig.
\ref{band_dos}(b), and discussed below] one also needs to
consider $U_p$ for these states along with $U_d$ for Zn-3$d$ states. 
The largest bandgap is obtained for $U_d =10$ eV and $U_p = 2$ eV.  
In this context it should be noted that application of $U$-term correlations
simultaneously to cation $d$ and anion $p$ 
states is not unprecedented, the most relevant example being ZnO
\cite{bashyal2018,oba2018}. At the same time, a similar large value of
$U_d = 10$ eV was also suggested in Ref. \cite{ma2013}.

The atomic contributions to
DOS across the Fermi energy within GGA+$U$ remain remarkably similar to the GGA
results, and is shown in Fig. \ref{band_dos}(b).
The valence band region upto $-5$ eV is mainly composed of $p$ states, with
dominant contribution by P-$3p$, followed by 
Sn-$5p$ and Zn-$4p$. The valence band maximum is composed primarily of
the P-$3p$ states, and the Zn-$4s$ states lie relatively deep in the
valence band, between $-4$ and $-5$ eV. 
%The Zn-$3d$ states lie
%relatively localized between $-5.5$eV and $-7.6$ eV.} 
On the other hand, the conduction band region is contributed by the P-3$p$,
Sn-$5s$ and $5p$ states, reflecting strong covalency effects in CH-ZSP [see Fig. \ref{band_dos}(b)].

The most notable difference between GGA and GGA+$U$ is the relative
position and spread of the Zn-$3d$ states.  Arguably, the Zn-3$d$ states within
GGA are over-hybridized with the Zn-$4p$ states similar to other strongly
covalent systems involving Zn \cite{oba2018}. As a result, they are underbound
and lie somewhat higher in the energy, in the range of -5.1 eV to -7.6 eV, (just) below the P-$4p$ states
in the valence band. This, in turn, leads to severe underestimation of bandgap within GGA \cite{bashyal2018,oba2018}. 
The location of the Zn-3$d$ states can, in principle, be tuned within
the GGA+$U$ functional.  
Within $+U$, they shift somewhat lower in energy and are more localized
(smaller bandwidth), leading to a well-defined gap in the DOS at $\sim -5$ eV. This
is accompanied by redistribution of the Zn-$4s$ and Zn-$4p$ contributions. 
Eventually, the $+U$ method improves
the bandgap over GGA, but is not sufficient. 

On the other hand, application of the TB-mBJ potential leads to a gap of $\sim 1.4$
eV [see Fig. \ref{band_dos}(c)] in good agreement with previously
reported value \cite{xu2015}. 
This is a significant improvement over the GGA and GGA+$U$
values, however remains only at $\sim 83 \%$ of the experimentally reported value. 
This is not surprising since P-$3p$ states contribute significantly to the
states across the Fermi energy \cite{koller2011}. 
Such a discrepancy is indicative of the fact that many body effects could be
important for CH-ZSP and that an approach accounting for such many
body effects, such as DFT calculations with hybrid functional or GW
approximation, may be required to address the full bandgap issue here.  

To compare the TB-mBJ bandstructure with that of GGA, a scissor-shift
$\Delta = 0.71$ eV is required, as shown in the lower panel of  Fig. \ref{band_dos}(c). 
However, a somewhat smaller scissor-shift of $\Delta=0.45$
shows a remarkably good qualitative and quantitative agreement between the two
methods [see the top panel in Fig. \ref{band_dos}(c)]. 
Comparison of atom-resolved DOS (not shown) suggests that the
relative compositions of the conduction bands is similar in both
approaches. 
Therefore, within all the considered approaches, the valence band
edge remains largely unaffected while the qualitative description of the
conduction bands and it's composition is nearly the same.

To summarize, within the considered methods,
the description of CH-ZSP predominantly differs only in the predicted value of the bandgap. Therefore,
    different values of scissor-shift is required to compare the GGA
    results with the others. A large value of $\Delta
    \sim 1$ eV is needed to match with the experimental results, whereas $\Delta = 0.36$ eV and $\Delta = 0.71$ eV is
    needed to compare with the GGA+$U$ and the TB-mBJ results,
    respectively. 

\begin{figure}[b!]
\includegraphics[width=0.485\textwidth,angle=0]{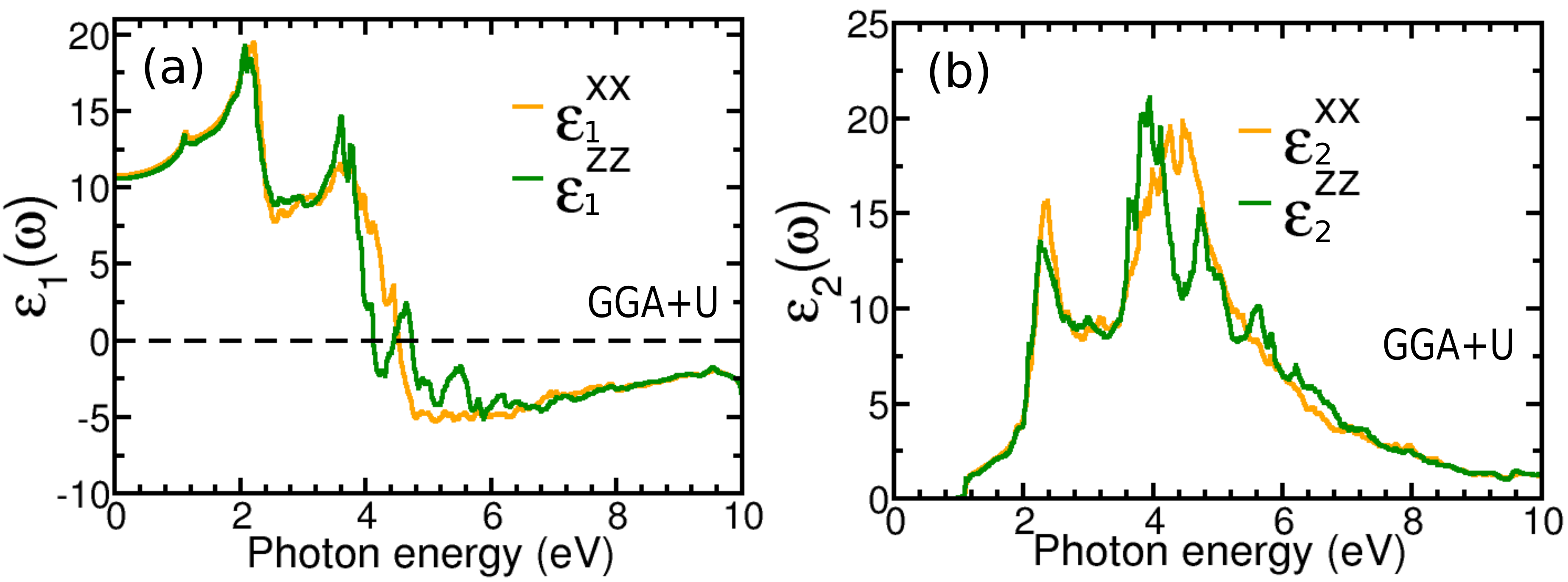} 
\includegraphics[width=0.195\textwidth,angle=-90]{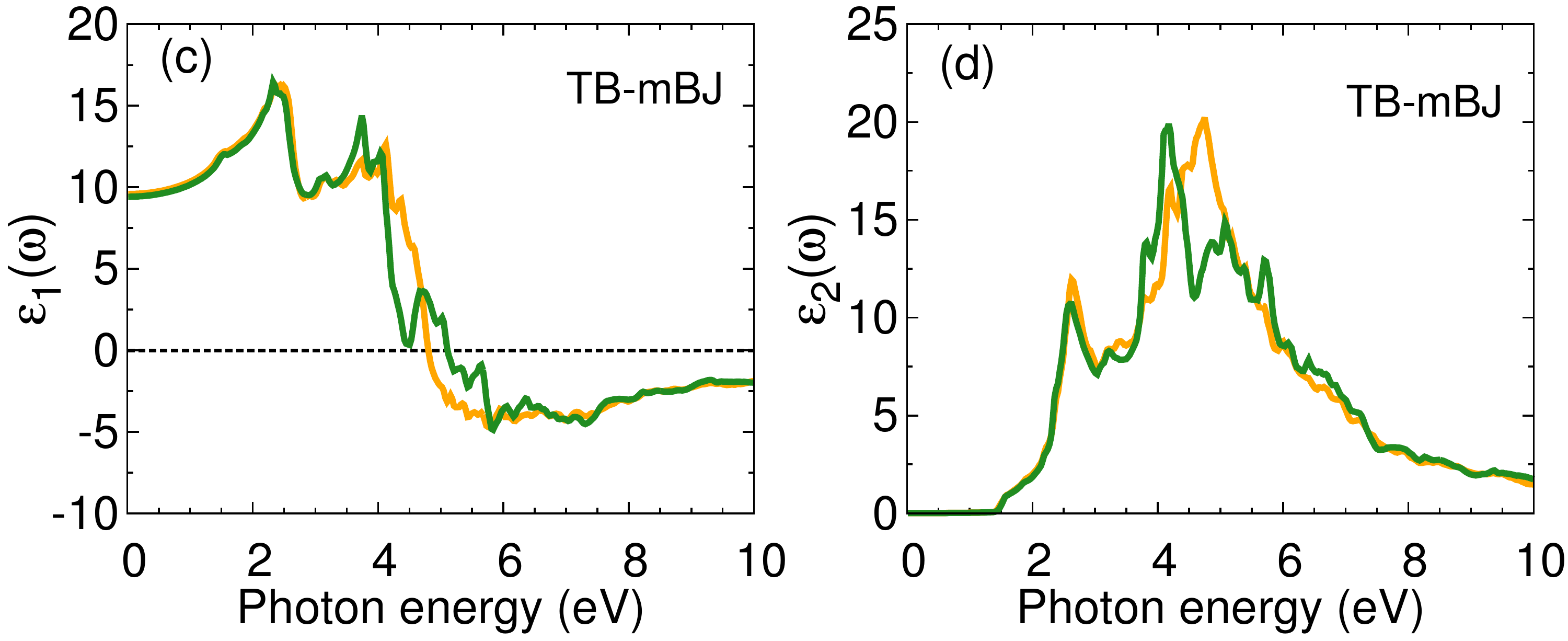} 
\caption{The real and imaginary parts of the dielectric constant
    as a function of the incident photon energy obtained within (a)-(b)
    GGA+$U$, and (c)-(d) TB-mBJ scheme, showing overall agreement.}
\label{epsa} 
\end{figure}

To further ascertain the degree of agreement between the
    considered methods, we also compare the linear optical response.
Fig. \ref{epsa} shows the
real and imaginary part of dielectric function obtained within
GGA+$U$ and it's comparison with the corresponding results from TB-mBJ. Tetragonal symmetry
of the crystal structure implies that the in-plane ($\alpha\beta = xx$,$yy$) and
the out-of-plane ($\alpha\beta = zz$) components are distinct. 
The qualitative similarities in both the schemes is evident. Within
GGA+$U$, the real
part of the dielectric constant $\epsilon_1$ has prominent peaks at
approximately 2 eV and 3.7 eV, and the zero-energy crossings lie between 
4.5 - 5 eV. 
The imaginary part of the dielectric constant $\epsilon_2$
is also characterized by two prominent peaks,
at $\sim 2$ eV and between 4.0-4.5 eV, similar to $\epsilon_1$.
These peak positions correspond to the interband
transitions between the valence and conduction band states. The dominant peak in
$\epsilon_2^{zz}$ (at $\sim 4$ eV) lies slightly lower than in $\epsilon_2^{xx}$, as expected
from the respective zero-energy crossings in $\epsilon_1$.
Considering a scissor-shift of $\Delta \sim 0.35$, required to match the bandgaps
between GGA+$U$ and TB-mBJ methods, the peak position in $\epsilon_1$ and 
$\epsilon_2$, as well the zero-energy crossing in $\epsilon_1$ are in
good agreement within the two approaches.

The zero frequency limit of $\epsilon_1(\omega)$, 
$\epsilon_1(0)$, is an important
quantity. It represents the electronic part of the static dielectric
constant and depends strongly on the bandgap. The static dielectric
constant is found to be $\epsilon_1(0) = 10.56$ eV within GGA+$U$.
In comparison, this value was measured to be 10 eV
\cite{madelung2012, petousis2017} while the TB-mBJ
calculations yield 9.7.

As the non-linear optical
properties depends crucially on the wavefunctions \cite{baltz1981}, 
reliable estimates of the optical properties, especially the magnitude
of the shift current, can thus be obtained even within the scissors
operator method GGA$+\Delta$. 
Therefore, in the following, we focus on the GGA$+\Delta$ and
GGA$+U$ methods with the understanding that an additional scissor-shift of $\sim
0.6$ eV ($\sim 0.4$ eV) may be required for a quantitative comparison with
experiments (TB-mBJ).

\begin{figure}[h!]
\includegraphics[width=0.48\textwidth,angle=0]{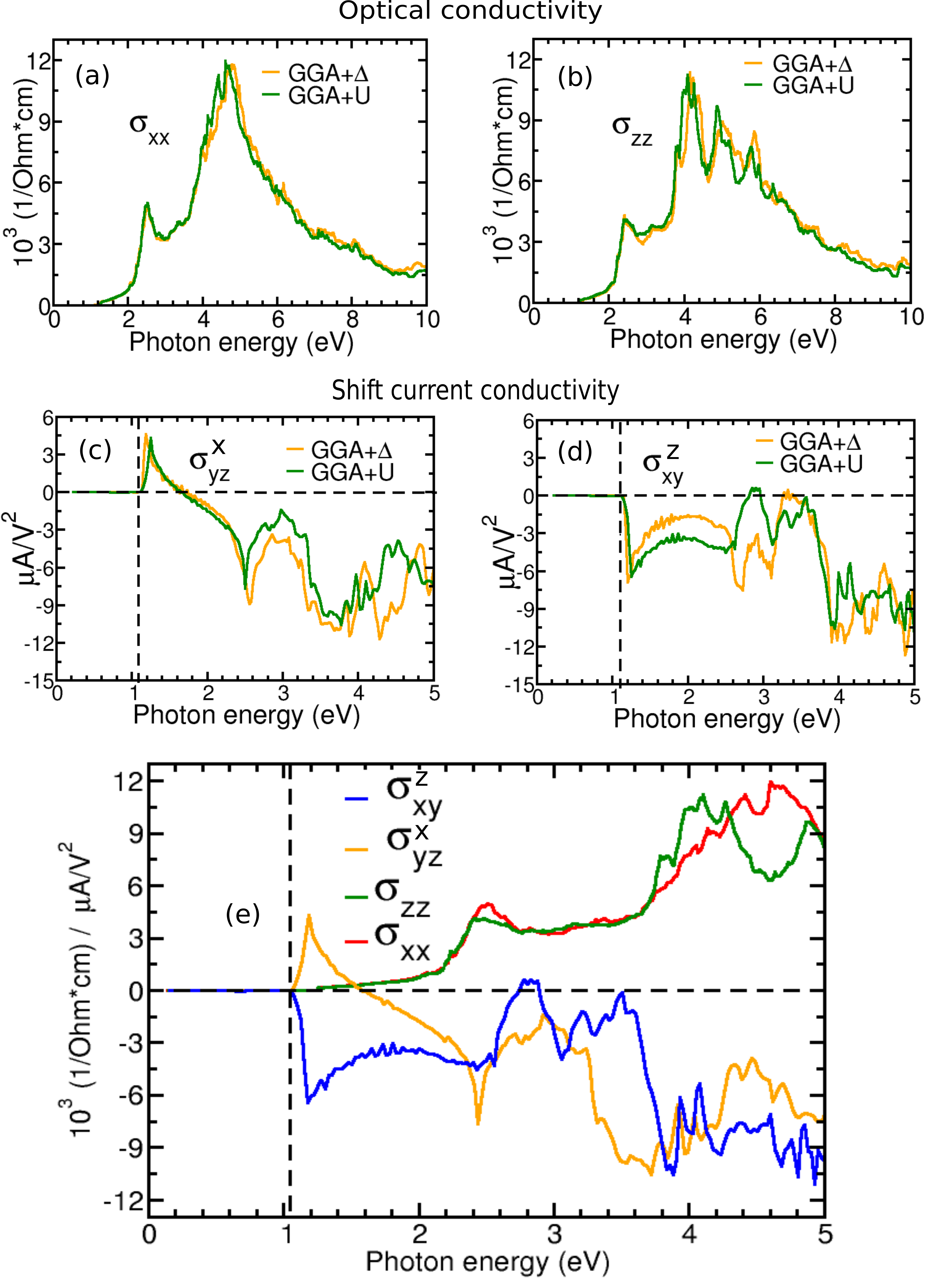} 
\caption{ The optical conductivity (a) $\sigma_{xx}$ and (b)
    $\sigma_{zz}$ from GGA+$U$ and GGA+$\Delta$ ($\Delta = 0.36$ eV). The
shift current conductivity (c) $\sigma{^x_{yz}}$ and (d)
$\sigma{^z_{xy}}$ from GGA+$U$ and GGA+$\Delta$. (e) Both the optical and
shift current conductivity from GGA+$U$ plotted on the same
scale.}
\label{sigma} 
\end{figure}

Figure \ref{sigma} shows the calculated optical and shift current
conductivity for CH-ZSP. The obtained GGA response has been scissor-shifted by
$\Delta = 0.36$ eV to compare with GGA+$U$. The structure and
magnitude of the optical response does not depend too much on the on
GGA+$U$.
We begin with a comparison of the optical conductivity obtained
within GGA and GGA+$U$, shown in Fig. \ref{sigma}(a) and \ref{sigma}(b), respectively, for
$\sigma_{xx}$ and $\sigma_{zz}$ . 
The optical conductivities
are in very good agreement, as expected from the fact
that both of these methods provide qualitatively similar description of the conduction
and valence bands. The main optical
conductivity peak in $\sigma_{xx}$ appears at 4.61 eV with a low energy
peak at 2.52 eV and for $\sigma_{zz}$, the peak appears at 4.08 and 2.43
eV, respectively. These peak positions are consistent with the
corresponding peak positions in the imaginary
part of the dielectric function $\epsilon_2$. 

In the shift current conductivity [see Fig. \ref{sigma} (c - e)],
$\sigma{^x_{yz}}$ and $\sigma{^z_{xy}}$ are the only nonvanishing,
independent components of the third rank tensor $\sigma{^\gamma_{\alpha
\beta}}$.  Similar to the optical conductivity, the shift current
response starts only above the bandgap. Interestingly, the
shift current shows a strong increase at the gap edge in contrast to the optical
current conductivity, which increases slowly above the gap. Figure
\ref{sigma}(c-d) shows the calculated shift current for
$\sigma{^x_{yz}}$ and $\sigma{^z_{xy}}$ components. The shift current
response for both ${xy}$ and ${yz}$ polarized light are negative though
$\sigma{^x_{yz}}$ has small positive contribution near the bandgap. The
shift current conductivity is around 6 ${{\mu}A}/V^2$ near the bandgap
for both $\sigma{^x_{yz}}$ and $\sigma{^z_{xy}}$ which is comparable to
recent experimental observations on the semiconductor SbSI
\cite{sotome2018}, and an order of magnitude larger than the famous
multi-ferroelectric compound bismuth ferrite (0.5 ${{\mu}A}/V^2$)
\cite{young2012_bfo}. Similar to the optical conductivity discussed before,
the shift current exhibits a large increase to 12 ${{\mu}A}/V^2$ at
photon energy at ($\sim 3.5-4$) eV. This is due to the large real space charge
center shift between valence electrons and conduction electrons, which
contributes mainly from $3p$ orbitals of P atoms to $5p$ orbitals of Sn
atoms.

%%%%%%%%%%%%%%%%%%%%%%%%%%%%%%%%%%%%%%%%%%%%%%%%%%%%%%%%%%%%%%%%%%%%%%%%%%%%%%%%%%%%%%%%%%%%%%%%%%%%%%%%%%%

\section{Conclusion and outlook}

In conclusion, we investigated the non-linear photocurrent in
non-centrosymmetric chalcopyrite semiconductor ZnSnP$_2$ based on first
principles calculations. Based on a detailed analysis of the electronic
properties of CH-ZSP within the traditional scissors operator method
GGA+$\Delta$,
GGA+$U$ and TB-mBJ methods, we find that TB-mBJ leads to a much better
agreement ($\sim 83\%$) with the reported experimental bandgap. More
importantly, although various methods rely on different approaches, the
description of the electronic bands within all these methods is
remarkably similar. This bodes well for the reliability of our estimates for linear and
non-linear optical properties based on either of the methods.
The shift current conductivity that we find is around 6 ${{\mu}A}/V^2$
near the bandgap and 12 ${{\mu}A}/V^2$ at photon energy $3.5-4\,$eV.
This comes mainly from the large real space charge center shift between
valence electrons and conduction electrons of P-3p and Sn-5p orbitals.
Distinct from the diffusion mechanism in the p-n junction based
photogalvanic effect, the generation of photocurrent under linear
polarized electromagnetic radiation in ZnSn$P_2$ is dominated by Berry
phase related shift current. Due to the underlying selection rules
ultrafast photo-induced currents will strongly depend on the crystal
orientation and laser polarization. This can offer a promising avenue to
achieve efficient generation and control of secondary terahertz
radiation, which in ZnSn$P_2$ will result from the intrinsic shift
current mechanism \cite{gao2020}: the magnitude of the shift current is
comparable to the recent experimental value on SbSI \cite{sotome2018}
and an order of magnitude larger than multi-ferroelectric compound
bismuth ferrite (0.5 ${{\mu}A}/V^2$) \cite{young2012_bfo}.

%%%%%%%%%%%%%%%%%%%%%%%%%%%%%%%%%%%%%%%%%%%%%%%%%%%%%%%%%%%%%%%%%%%%%%%%%%%%%%%%%%%%%%%%%%%%%%%%%%%%%%%%%%%
\section*{Acknowledgements}
We thank Manuel Richter for helpful discussions and Ulrike Nitzsche for
technical assistance. This work was supported by the German Research
Foundation (DFG) via SFB 1143, Project No. A5 and  by the DFG through the
W\"urzburg-Dresden Cluster of Excellence on Complexity and Topology in Quantum
Matter - {\it ct.qmat} (EXC 2147, Project No. 39085490). 

%%%%%%%%%%%%%%%%%%%%%%%%%%%%%%%%%%%%%%%%%%%%%%%%%%%%%%%%%%%%%%%%%%%%%%%%%%%%%%%%%%%%%%%%%%%%%%%%%%%%%%%%%%%

\appendix

\section{Role of spin-orbit coupling (SOC)}
\label{sec:app_a}

\begin{figure}[h!]
    \centering
    \includegraphics[angle=0,scale=0.28]{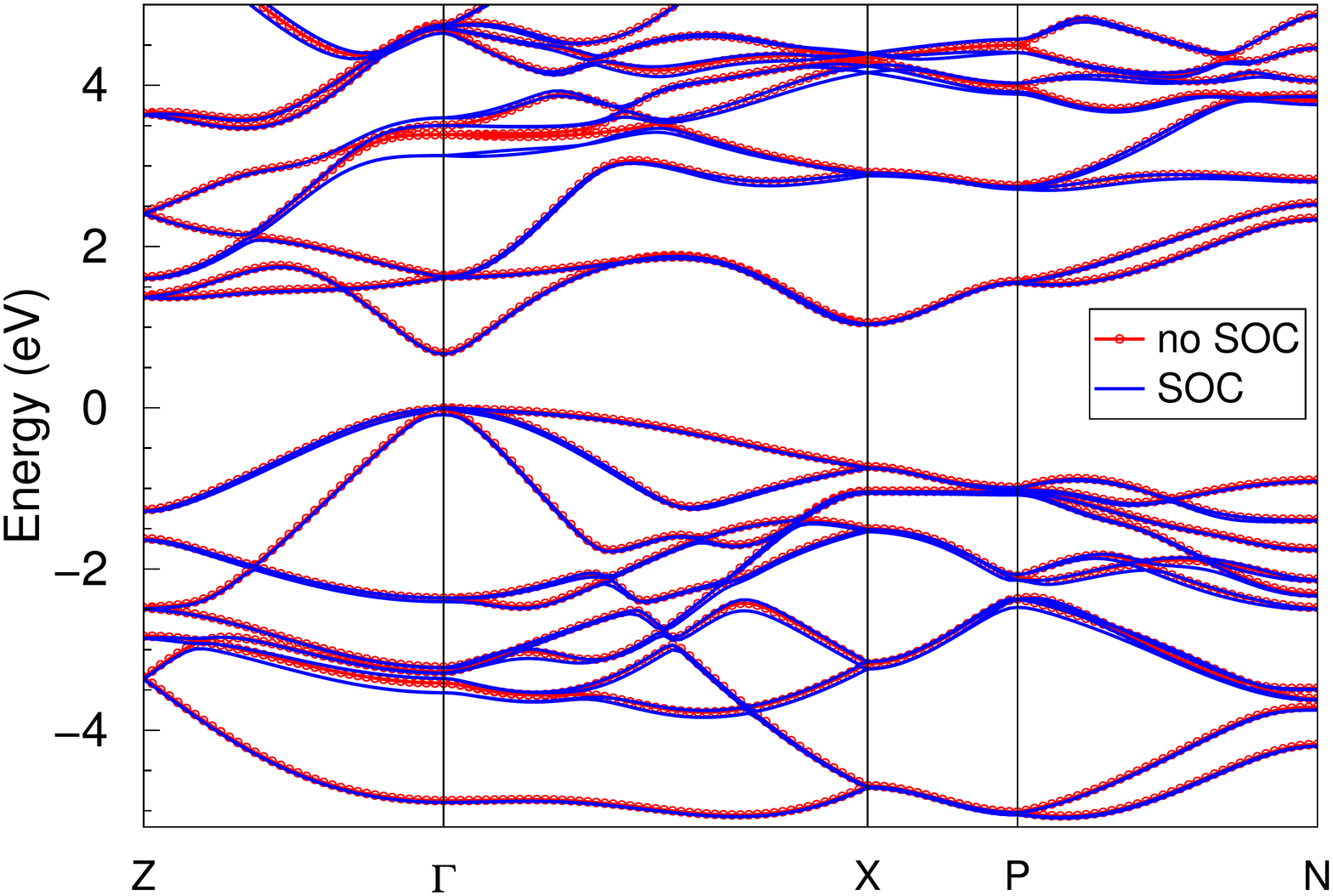}

    \caption{A comparison of the scalar relativistic (no SOC) and
        full relativistic (SOC) bandstructures for ZnSnP$_2$.
    }
    \label{fig:bands_so}
\end{figure}

Figure \ref{fig:bands_so} shows a comparison of the bandstructures of
CH-ZSP within the scalar relativistic (``no SOC") and full relativistic
GGA calculations. Sizable differences are found only around $-3\,$eV and $3.5\,$eV, where
the Sn-$5p$ contribution is dominant (see Fig. \ref{band_dos}). 

\bibliography{zsp}

%%%%%%%%%%%%%%%%%%%%%%%%%%%%%%%%%%%%%%%%%%%%%%%%%%%%%%%%%%%%%%%%%%%%%%%%%%%%%%%%%%%%%%%%%%%%%%%%%%%%%%%%%%%

\end{document}